# Vertical Displacement Events in Shaped Tokamaks

A. Y. Aydemir
*Institute for Fusion Studies*
*The University of Texas at Austin*
*Austin, Texas 78712 USA*

## Abstract

Computational studies of vertical displacement events (VDE's) in shaped tokamaks are presented. The calculations are performed with our nonlinear, 3D resistive MHD code, **CTD**, which can efficiently treat resistive walls and free boundary displacements in moderately-shaped geometries. This work has a number of related goals: First, the mechanisms for generation of halo currents, the paths taken by them in the plasma and surrounding conductors, and their relative magnitude, are elucidated. Second, coupling between an $n = 0$ vertical instability and an $n = 1$ external kink mode are examined to offer a possible explanation for the nonuniformities observed in the poloidal halo currents during VDE's, and the forces generated by them on plasma-facing components. Finally, effects of a rotating liquid metal wall on the equilibrium and $n = 0$ stability of elongated plasmas are briefly discussed.

## I. INTRODUCTION

Major disruptions, sudden and usually unanticipated loss of confinement, may be an unavoidable feature of tokamaks. These events are characterized by a rapid loss of thermal energy, the "thermal quench", which typically occurs in less than a millisecond, followed by the current quench and the transfer of a large fraction of the plasma current to the surrounding conducting structures, either inductively, or through direct coupling. Since complete avoidance of disruptions may not be possible, mitigating their damaging effects through a better understanding of the various events that accompany them is a research goal of many experimental and theoretical efforts.

Vertical displacement events (VDE's) are usually a consequence of a disruption. The drop in the internal energy during a disruption is accompanied by a redistribution of the current and a toroidally inward shift of the plasma column. At this new toroidal location, ratio of the quadrapole field, used for poloidal shaping, to the dipole field, which helps provide toroidal equilibrium, tends to be such that the plasma becomes susceptible to an axisymmetric vertical instability. The nonlinear evolution of this instability eventually brings the plasma in contact with the vacuum vessel or other structures. On other occasions, a VDE is the cause of the disruption. Here, the vertical motion of the column, either due to an inherent instability, or failure of the vertical position control system, precedes the thermal quench, which now occurs as the plasma comes in contact with the vessel. Both types of events are well documented and observed on all major tokamaks[1–4]. They are also a source of major concern for future devices[5]. A particularly detailed discussion of the experimental results, from Alcator C-Mod, is presented by Granetz, *et al.*[6]. A recent and useful review from a theoretical point of view of various disruptive events in tokamaks and the related terminology can be found in Humphreys and Kellman[7].

In this work, a vertical displacement event (VDE) will be defined to include any uncontrolled vertical motion of the plasma column in tokamaks that brings it in contact with the surrounding structures. Thus, it will encompass those displacements that are preceded by a pre-disruption thermal quench, as well as those caused by a loss of vertical control that leads to a disruption.

In addition to the possibly damaging heat loads it produces at the plasma-vessel contact points, a VDE can lead to dangerous electromechanical stresses because of the currents it generates in plasma-facing structures. These currents and the associated mechanical stresses, rather than the induced the thermal stresses, will be one of the main topics of this work. If the poloidal currents in the vessel, usually referred to as the "halo currents", are toroidally uniform, their interaction with the toroidal field merely produces a net vertical force on the vessel. If the toroidal distribution is nonuniform, as it seems to be the case experimentally[6] then the resulting forces lead to a net torque that needs to be taken into account when considering the stresses on the vessel. This toroidal nonuniformity in the poloidal halo currents can be due to the interaction of the $n = 0$ vertical mode with an $n = 1$ external kink and will be second major topic of discussion in this work.

The outline of this paper is as follows: The next section will briefly summarize the computational model used in this work. Section III will examine axisymmetric VDE's and the induced eddy and halo currents in the vessel. In Section IV, the axisymmetry constraint will be removed and the evolution of VDE's in the presence of kink modes that lead to toroidal asymmetries will be discussed. Finally, Section V will briefly examine the effects of flowing liquid metal walls on MHD equilibrium and stability.

## II. THE COMPUTATIONAL MODEL

Our three dimensional, toroidal, nonlinear MHD code, **CTD**[8,9], has been modified to include a vacuum region and a resistive wall to be able to address, not only the three dimensional VDE problem that will be considered here, but nonlinear external kink modes in general. The recent extension of the resistive wall to include arbitrary fluid motions makes it possible to study the role of liquid metal walls in controlling MHD modes, as will be discussed briefly at the end of this article. The equations solved by **CTD**, written in nondimensional form, are

$$\frac{\partial \rho}{\partial t} + \nabla \cdot (\rho \mathbf{u}) = 0 \tag{1}$$

$$\rho \left\{ \frac{\partial \mathbf{u}}{\partial t} + \mathbf{u} \cdot \nabla \mathbf{u} \right\} = \mathbf{J} \times \mathbf{B} - \nabla p + \mu_i \nabla^2 \mathbf{u}, \tag{2}$$

$$\frac{\partial \mathbf{A}}{\partial t} = \mathbf{u} \times \mathbf{B} - \eta \mathbf{J} - \nabla \times \mu \nabla \times \mathbf{J}, \tag{3}$$

$$\frac{\partial p}{\partial t} + \mathbf{u} \cdot \nabla p = -\Gamma p \nabla \cdot \mathbf{u} + \kappa_\parallel \nabla_\parallel^2 p + \nabla \cdot (\kappa_\perp \nabla p) + (\Gamma - 1)\eta J^2. \tag{4}$$

The Lundquist number (the magnetic Reynolds number) $S$, ratio of the resistive diffusion time to the poloidal Alfvén time, is given by $S = 1/\eta(0)$, where $\eta(0)$ is the value of resistivity on axis. Here the resistivity is a three dimensional function of temperature and possibly other fields: $\eta = \eta(p(\mathbf{r},t),\mathbf{r},t)$. $\kappa_\parallel$ and $\kappa_\perp$ are the parallel and perpendicular thermal conductivities, respectively, and $\Gamma$ is the ratio of specific heats. The electron viscosity (hyper-resistivity) term in the Ohm's law, which was used in a previous study examining fast sawtooth crashes[10], will not be used in this work ($\mu = 0$).

Noncircular geometries are treated using a conformal transform[11,12,10] from the configuration space to a unit circle in the computational domain. This technique has proven to be highly efficient compared to alternatives when the boundary deformation is not too severe.

The equilibria discussed here have have been calculated self-consistently with **CTD** in the presence of various external currents that provide shaping and toroidal equilibrium, a double or single-null field geometry, and various transport processes.

The next section describes studies of axisymmetric VDE's using **CTD**.

## III. AXISYMMETRIC VDE'S

This and the following sections present salient features of "numerical experiments" performed with **CTD** on vertical displacement events. These numerical experiments differ from the laboratory VDE's in a number of ways, most important of which is the compressed timescale in our "experiments" due to numerical limitations on transport coefficients such as the resistivity; we typically use a magnetic Reynolds number of $S \sim 10^5 - 10^6$, a value that is approximately three orders of magnitude lower than what one would see in a modern tokamak. Also, a self-consistent study of the thermal quench and the MHD activity that leads to it are computationally very time consuming and beyond the scope of this work. Thus, this phase of the disruption is replaced by an artificially induced thermal quench during which a significant portion of the internal energy is radiatively damped away. This artificial radiation cooling may in fact be not unphysical, since there is some evidence[13] that this mechanism dominates over the anomalous thermal conduction induced by nonlinear MHD during the latter stages of disruptions. Finally, plasma-wall interaction in our single-fluid resistive MHD model necessarily leaves out a large number of complicated issues which are also beyond the scope of this work.

Typical evolution of an axisymmetric VDE in our numerical experiments is shown in Figure 1; Fig. 1a depicts the poloidal flux contours at various times as the plasma moves down and hits the resistive wall, shown by the thin layer on the outside between two boundary lines.

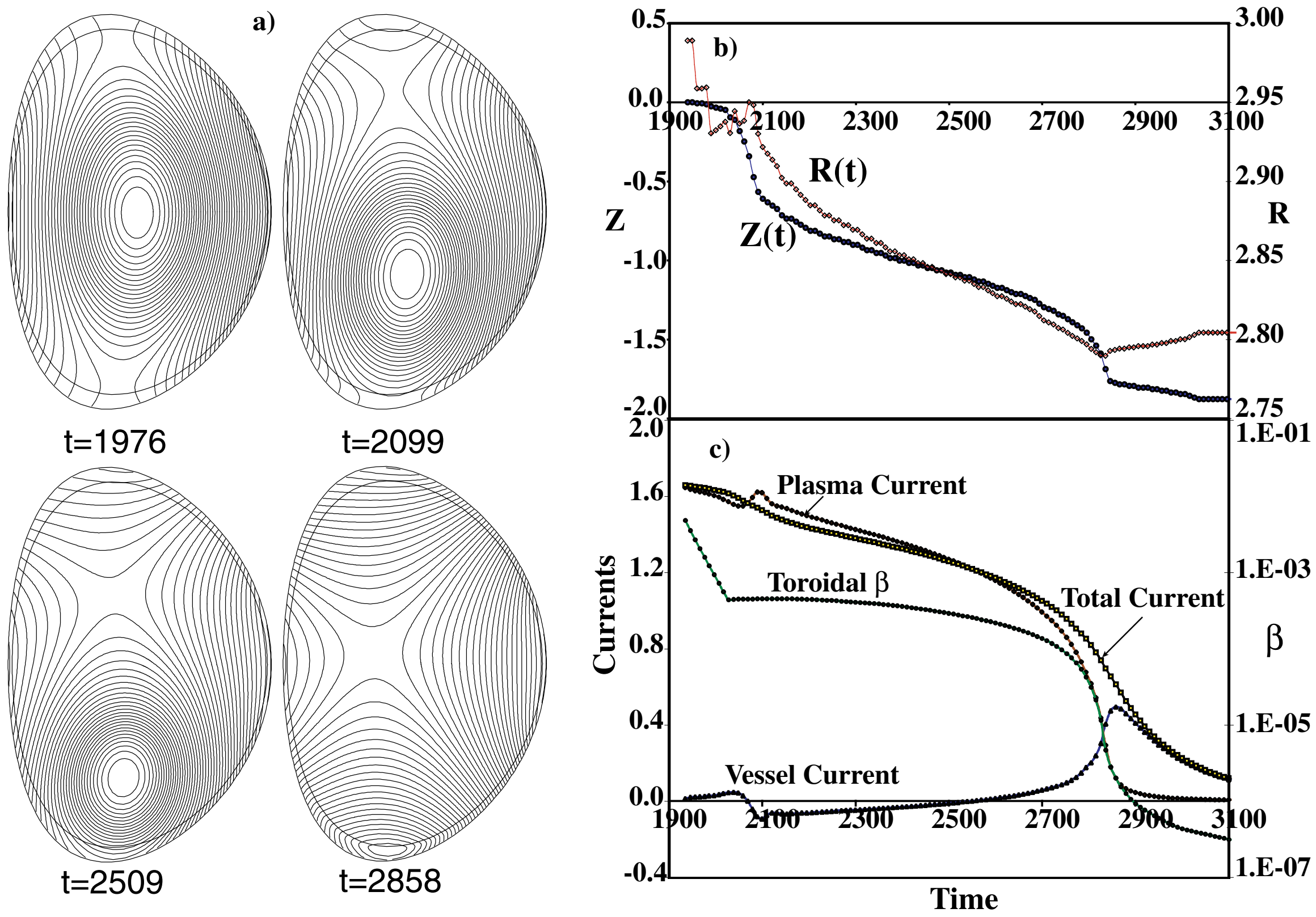

FIG. 1: a) Time evolution of poloidal flux during an axisymmetric VDE. b) Radial and vertical location of the magnetic axis during the VDE. c) Time evolution of toroidal currents and $\beta$.

Figure 1b shows time evolution of the magnetic axis coordinates, whereas Fig.1c shows the plasma current, toroidal vessel current (toroidal current induced in the vacuum vessel), and toroidal $\beta$ for this VDE, which was initiated by a thermal quench in an equilibrium with an initial $\beta$ of 0.5%.

Here the toroidal $\beta$ is reduced to 10% of its initial value (a small but arbitrary number) as the internal energy is radiatively damped, after which it remains approximately constant until the plasma gradually comes in contact with the wall. It essentially drops to zero after $t \simeq 2800$ as the magnetic axis moves into the wall.

Before moving on to halo currents, first we examine some interesting temporal and spatial features of the toroidal currents, which will also help us with our understanding of the poloidal halo currents below.

## A. Toroidal currents

As it is seen in Fig. 1c, at $t \simeq 2100$, soon after the "thermal quench", the plasma current exhibits a "spike", the classic signature of a disruption (along with the negative spike in the loop voltage). In laboratory experiments, the positive jump in the current and the negative one in the voltage are attributed to an expansion of the current and expulsion of poloidal flux through the boundary[14,15]. The mechanism in our numerical experiments is conceptually related but necessarily different in the details, since we do not include in these calculations an MHD-induced expansion of the plasma current. Here it involves the toroidal currents generated in the scrape-off layer (SOL) and in the vacuum vessel during the VDE. Since this mechanism may be relevant to a better understanding of the laboratory experiments, we will discuss it in some detail.

The thermal quench by itself only leads to a shrinking of the current channel and cannot explain the current spike we observe. In fact, in calculations in which the plasma is allowed to go through a thermal quench but artificially prevented from a VDE, we observe no spike but only a gradual decay of the current. Thus, the increase in the plasma current during the VDE is associated with the dynamics of the VDE itself. The actual mechanism has two related components.

First, the VDE displacement is not that of a rigid body; as the core moves vertically, the outer layers gradually peal away and move initially laterally, then in the opposite direction to the VDE, providing a "return flow", as seen in the first frame in Fig. 2. This reverse return flow is not completely field-aligned and leads to an effective toroidal electric field through the second term in $\mathbf{E_{eff}} = \mathbf{E} + \mathbf{v} \times \mathbf{B}$, which drives a positive toroidal current localized around the separatrix as it moves down with the VDE.

Note that the decaying plasma current also induces a positive electric field (the first term on the right hand side of the above equation), which is not localized spatially and also is much smaller than the $\mathbf{v} \times \mathbf{B}$ term. Second part of the mechanism is the negative toroidal current induced in front of the moving plasma column (and in the wall) that serves to slow down the displacement. Both the positive toroidal current in the SOL and the negative current in front of the dropping column can be seen in the last three frames in Fig. 2. Note that the toroidal currents in the SOL are maximum during the fastest part of the VDE around $t \leq 2100$.

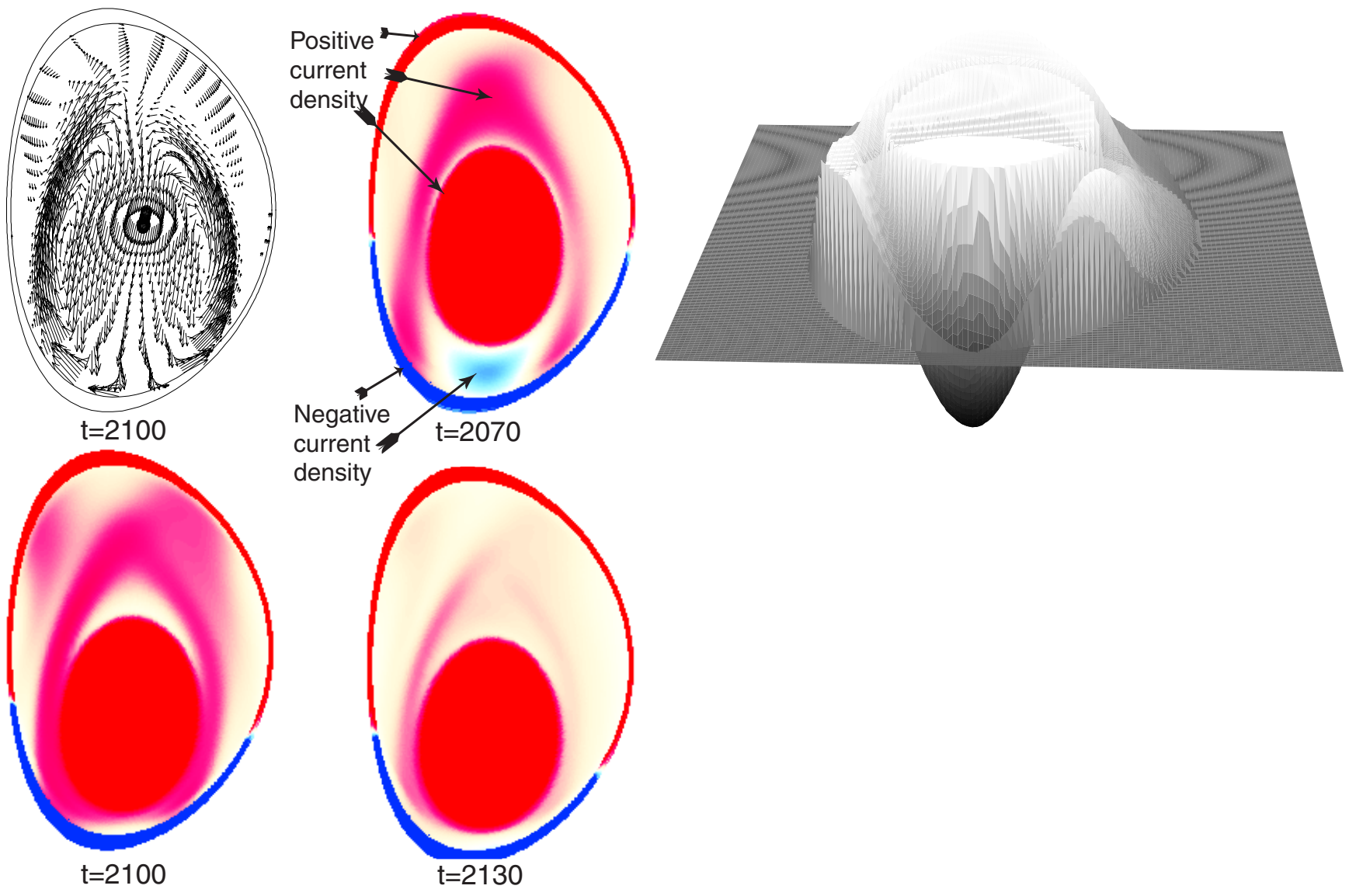

FIG. 2: First frame shows the fluid velocity vectors during the fastest part of the VDE. Note the return flows that drive a toroidal current around the separatrix. The remaining frames show the toroidal current density during the formation of the current spike. The vertical scale has been exaggerated to bring out the halo region more clearly. In the 3D figure, which shows the currents at $t = 2070$, the wall currents have been left out for clarity.

Note also the dipole nature of the current in the wall that helps slow down the VDE: the toroidal currents in the lower half of the wall are negative (in opposite direction to the main plasma current), and positive in the upper half. As the column approaches the wall, the negative current region leaves the plasma and enters the vessel, resulting in the positive jump in the plasma current and the negative one in the vessel current around $t \simeq 2100$ that we saw earlier in Fig. 1c. The total toroidal current in the plasma-wall system remains approximately constant during these rapid changes in its individual components, as can be seen in Fig. 1c. The time scale for changes in $I_T = I_p + I_v$, where $I_p$, and $I_v$ are the plasma and vessel currents respectively, is determined by the $L/R$ time of the vessel; in normalized units, we have, for this particular case, the vessel resistivity $\eta_v = 10^{-4}$, and the vessel thickness $\delta = 0.05$, which leads to an approximate vessel $L/R$ time of $\tau_v \simeq \delta a/\eta_v = 500$, which is much longer than the time scale of the current spike in $I_p$ we observe in Fig. 1c.

Next we consider the related subject of poloidal currents generated in the plasma and the vessel.

## B. Poloidal currents

Here we will distinguish between the poloidal currents in the plasma core, the region of closed field lines, and those on the open field lines of the SOL. In the literature, the vessel currents are often assumed to be extensions of the currents on the open field lines. However, we will show below that, at their maximum at least, the poloidal currents measured in the vessel are dominated by the former, *i.e.* the currents in the core plasma that couple to the wall, although the halo currents of the SOL do make an initial contribution.

Fig. 3 summarizes the temporal and spatial behavior of the poloidal vessel currents. In Figures 3(a,b), time variation of the radially and toroidally integrated poloidal current is shown as a function of the poloidal angle. In Fig. 3(c) the data at a single poloidal point near the bottom of the vessel where the current exhibits its maximum is shown. In this Figure, positive values have been chosen to correspond to paramagnetic currents.

Note that the start of the thermal quench is accompanied by a nearly symmetric diamagnetic poloidal current in the vessel, as seen in Figs. 3(b,c). This diamagnetic phase of the wall currents accounts for the large negative dip seen in Fig. 3(c) before $t \simeq 2100$. These currents are driven in response to the rapidly decreasing plasma pressure within the core region, leading to an increase in the toroidal flux, as the plasma pressure is replaced by the toroidal field. During this phase, the core becomes more paramagnetic if initally $\beta_p < 1$, or switches to a paramagnetic state if $\beta_p > 1$ before the quench. Note that force-free currents ($\nabla p = 0$) are necessarily paramagnetic.

As the VDE evolves, the diamagnetic currents in the wall are gradually replaced by currents in the paramagnetic direction (Fig. 3c), as the halo currents of the SOL, and the core paramagnetic currents, couple to the vessel. Below, we further examine the nature of these currents.

### *i. The currents in the SOL*

The poloidal electric fields due to the changing toroidal flux within the shrinking plasma cross-section undoubtedly makes a contribution to the halo currents in the SOL. However, a significant portion of these SOL halo currents are merely the poloidal projections of the force-free currents induced by the toroidal electric fields discussed in the previous

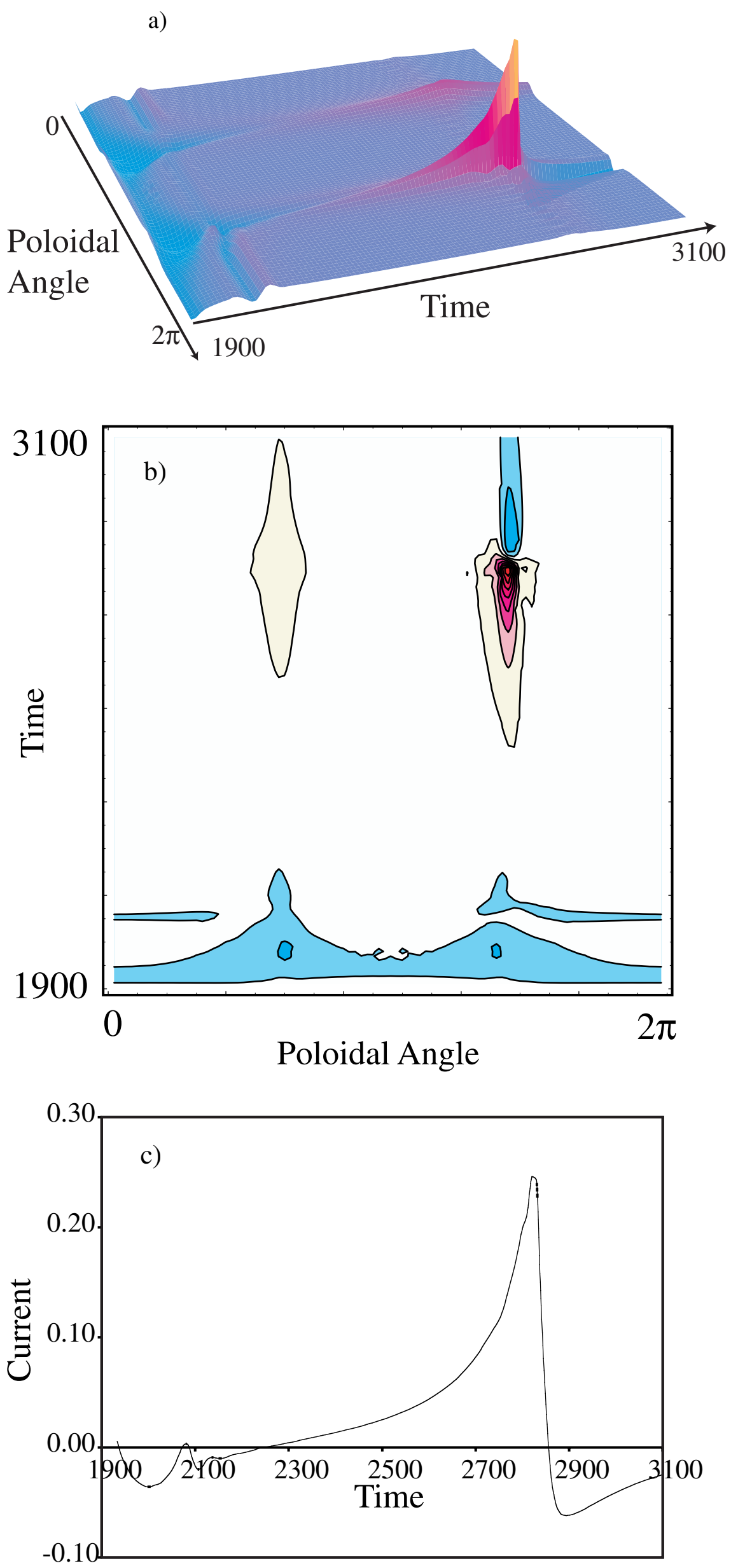

FIG. 3: Poloidal vessel currents. a,b) Currents as a function of poloidal angle and time. c) Currents at a single poloidal angle near the bottom X-point. Values shown are radial and toroidal integrals of the poloidal current density in the wall at each poloidal location. Positive values correspond to paramagnetic currents.

section (See Fig. 2).

Figure 4 shows the poloidal current vectors from the initial and final stages of a VDE; this particular numerical experiment had slightly different parameters than the one discussed above and was chosen because it shows more clearly the halo currents (Fig. 4a) that develop around the core plasma at the early stages of the VDE. The maximum wall currents occur, however, In Fig. 4b, sometime after the core currents directly couple to the vessel. In other words, the halo currents along the field lines that intersect the vessel (Fig. 4a) tend to form the earlier and smaller part of the vessel currents, and the sharp rise in the wall current and the maximum seen in Fig. 3c occur as the paramagnetic core currents start completing their paths through the vessel after plasma-wall contact.

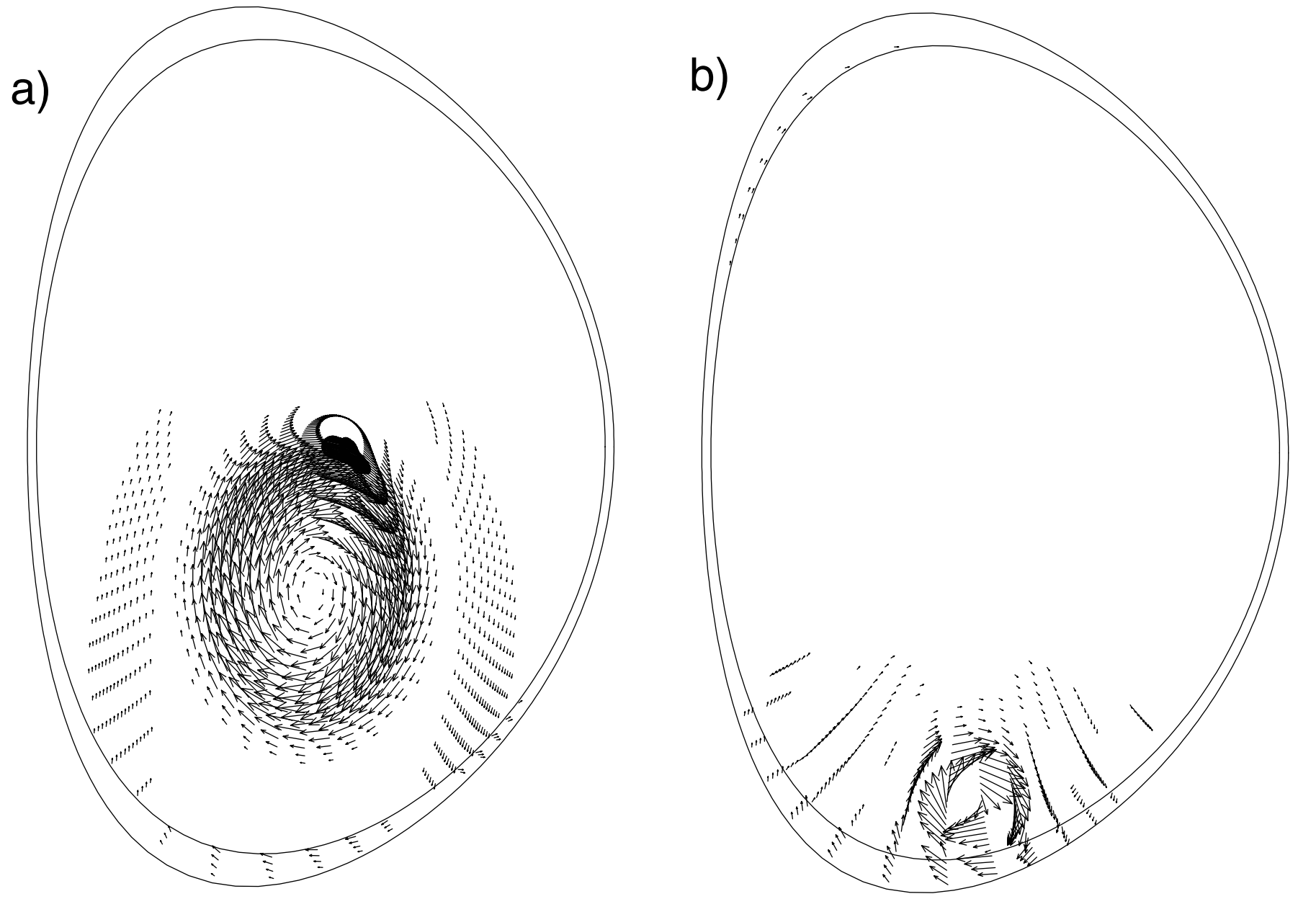

FIG. 4: a) Paramagnetic core plasma currents, and the halo currents around them during the initial stages of a VDE. b) Currents at the time of maximum poloidal vessel curents. Note that only the core plasma currents play a role here.

### *ii. The halo current maximum*

Since the vessel halo currents are dominated by the paramagnetic core plasma currents that couple to the vessel after plasma-wall contact, the maximum halo current possible, $I_{halo}^{max}$, is the total poloidal current in the plasma after the thermal collapse (when the plasma is force-free and paramagnetic). Using an axisymmetric flux-coordinate system $(\psi, \theta, \zeta)$, with $\zeta$ being the symmetry direction, we can easily show that the total poloidal current in the plasma is given by $I_\theta = -2\pi(I_1 - I_0)$, where $I = RB_{tor}$, $I_1 \equiv I(\psi_1)$, $I_0 \equiv I(0)$, and $\psi_1$ defines some appropriate boundary surface for the core, such as the one associated with $q_{95}$. Using the Grad-Shafranov equation (with $p' = 0$), $R^{-2}J^{\zeta} = \Delta^{\star}\psi = -II'$, and integrating for the total plasma current, we get, for monotonic $I'$,

$$I_p = -2\pi \int_0^{\psi_1} q \frac{dI}{d\psi} d\psi = -2\pi q_b (I_1 - I_0), \tag{5}$$

where $q_b \equiv q(\psi_b)$, and $\psi_b$ is some point in the interval $[0,\psi_1]$. These expressions for the poloidal and toroidal currents in the plasma then lead to $I_{halo}^{max} = I_\theta = I_p/q_b$. Assuming that the $q-$profile is uniform after the thermal quench with $q_b = q_{95}$, we get the expression $I_{halo}^{max} = I_p/q_{95}$. For monotonically increasing $q-$profiles, we have

$$I_{halo}^{max} = C \frac{I_p}{q_{95}}, \tag{6}$$

with the constant $C > 1$ being determined by the details of the post quench $q-$profile. This expression agrees with a number of experimental observations[6,16]. The large variation in the experimental data is probably due to the fact that the really relevant values for the parameters $I_p$, and $q_{95}$ are not their post-thermal quench values but the ones immediately prior to wall-contact, which may be quite different.

## IV. NON-AXISYMMETRIC VDE'S

As stated earlier, VDE's are not entirely axisymmetric ($n = 0$) events but tend to feature, to some degree, $n = 1$ and higher components. In laboratory experiments, the distribution of vessel halo currents show[6] a toroidal peaking factor (peak-to-average value) of $\sim 2$. Although there are other possible explanations for the asymmetry[17], obviously the interaction of an $n = 1$ kink mode with the $n = 0$ vertical instability during the VDE also offers a plausible explanation.

In this Section, we relax the axisymmetry constraint of the previous sections and follow nonlinear evolution of equilibria unstable to an $n = 1$ external kink mode that are also vertically unstable. In an actual VDE, the plasma presumably goes unstable to the kink after the start of the VDE. For numerical convenience, we start our experiments in a kink-unstable state.

For one of these equilibria, velocity vectors for the linear $n = 1$ mode and the eddy currents generated by it in the resistive wall are shown in Fig. 5. The mode is an $m = 2$, with a strong $m = 1$ component, although the equilibrium has $q_0 = 1.14$. Note that in this linear stage, the eddy currents are concentrated on the outboard side of the torus.

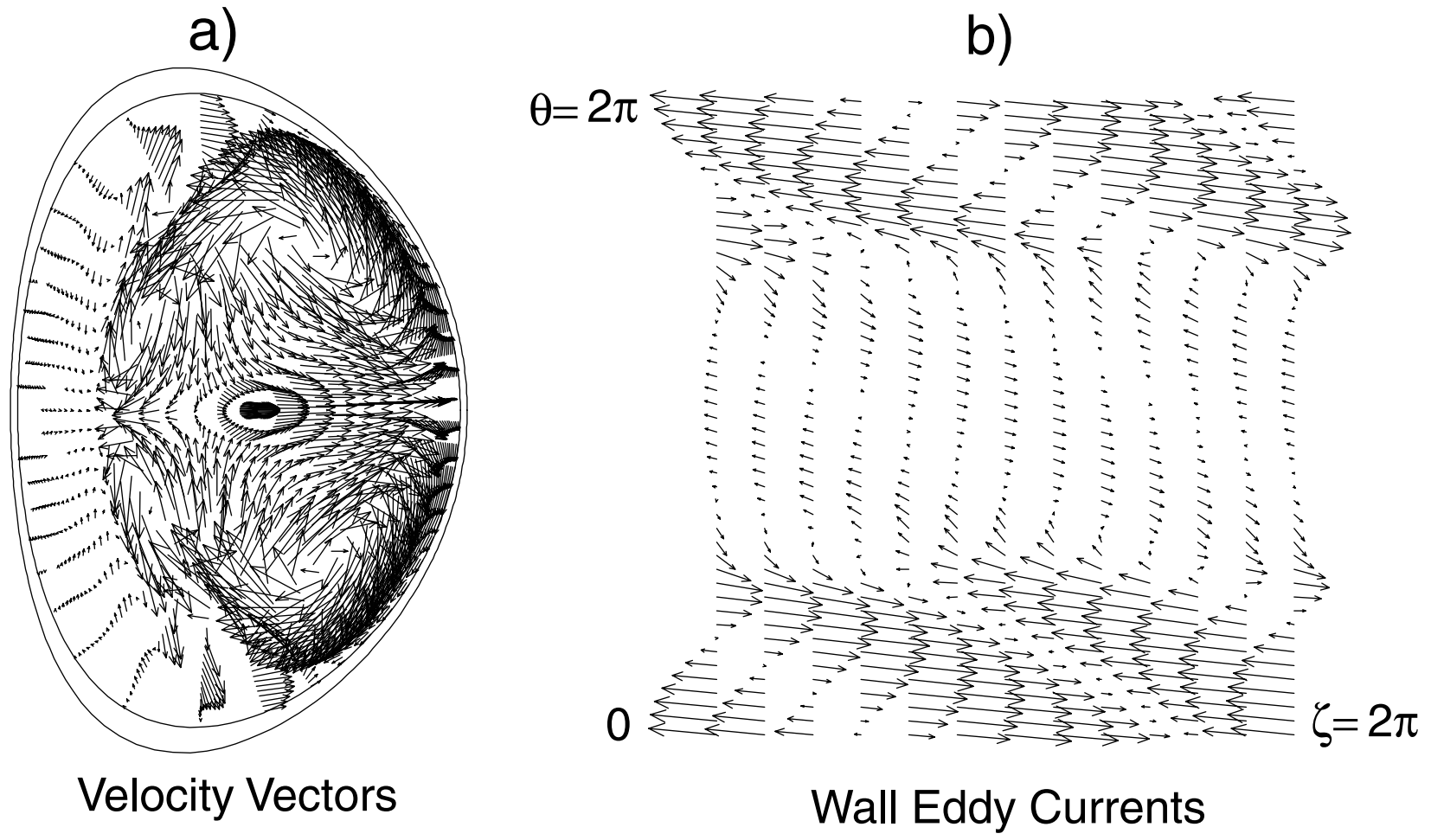

FIG. 5: a) Velocity vectors for the $n = 1$ eigenfunction. b) Eddy currents in the wall induced by the mode.

Although a study of these linear eddy currents may play a role in a future work involving feedback stabilization, our emphasis here is on the nonlinear effect of the $n = 1$ on the symmetric halo currents in the wall. Figure 6 summarizes this effect. Fig. 6a shows the (negative of the) halo currents in the wall at the time the current has reached its maximum. Fig. 6b shows the toroidal variation of the current at a fixed poloidal angle, whereas the last figure shows its poloidal variation for $\zeta = \pi$. The toroidal peaking factor for this case is approximately 1.6. Note the high degree of poloidal localization of the current near the bottom of the vessel, although some small current is observed near the upper X-point as a result of currents flowing along the open field lines that intersect the upper portion of the vessel.

## V. EFFECTS OF ROTATING LIQUID METAL WALLS

Use of a liquid metal blanket to deal with heat and neutron fluxes in fusion reactors is not a new idea[18]. It has been revived recently to address the high heat flux and MHD stability issues in advanced concept designs[19,20]. Since the $n = 0$ mode and the VDE's are prime candidates for stabilization by such a scheme, here we briefly examine the effects of rotating liquid metal walls on MHD eqilibirium and stability, leaving a more complete discussion of the subject to a future publication.

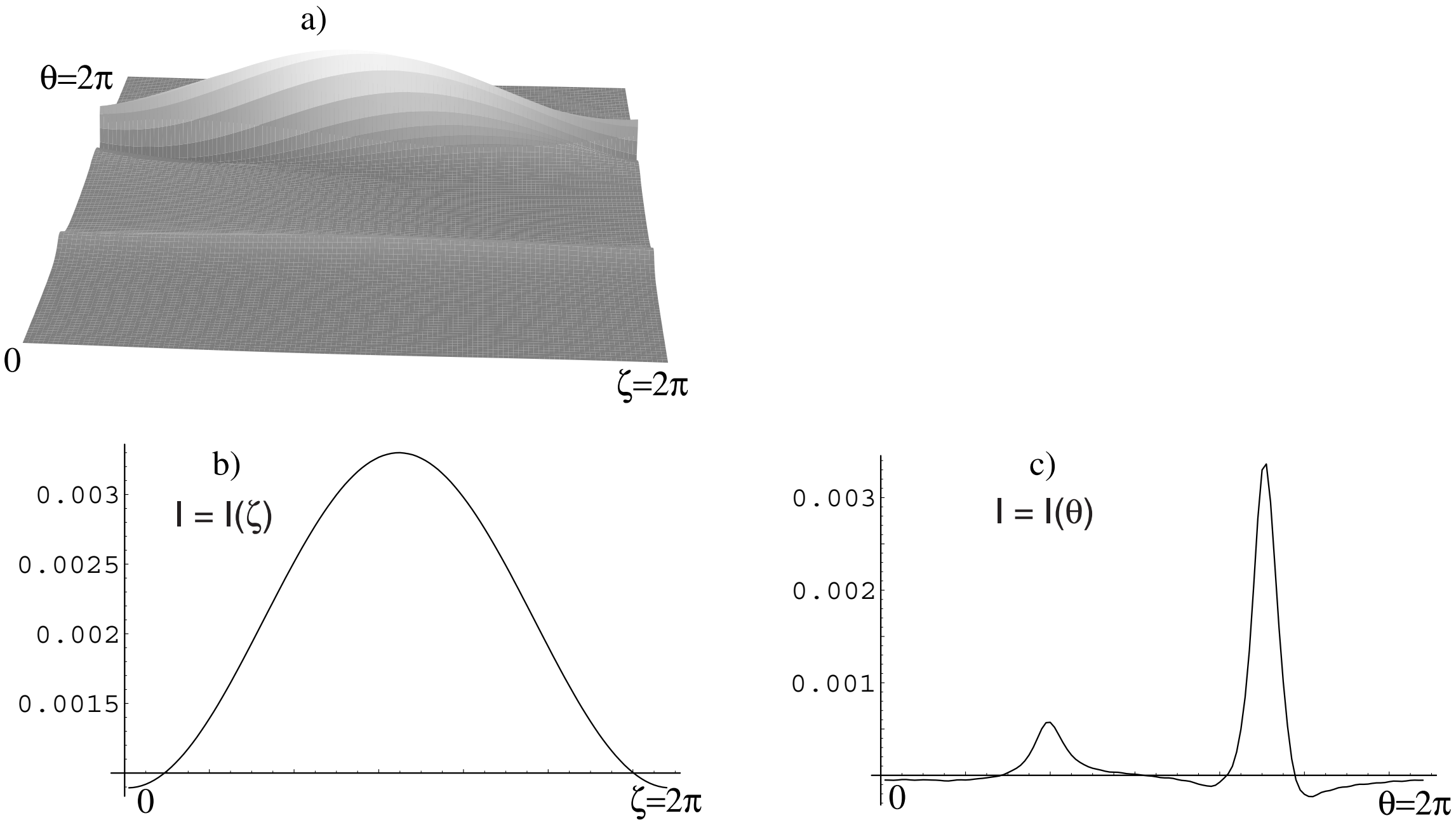

FIG. 6: a) Negative of the halo currents in the wall as a function of the poloidal and toroidal angles. b) Toroidal variation of the currents. c) Poloidal variation of the currents.

Desirability of a rotating liquid metal wall from the perspective of MHD stability is understood easily. A toroidal resistive wall of thickness $\delta$ and radius $a$ will have an $R/L$ time of $\tau_W = \delta a/\eta$; it will act as a low-pass filter and screen out frequencies higher than $\Omega_W = 2\pi/\tau_W$. If the wall is moving with velocity $\mathbf{u}$, then a mode with frequency and wavenumber $(\omega, \mathbf{k})$ in the lab frame will have a frequency $\Omega_R = \omega + \mathbf{k} \cdot \mathbf{u}$ in the rest frame of the wall. If $\Omega_R \gg \Omega_W$, the resistive wall will appear to be "perfectly conducting" and tend to stabilize the mode. In reality, this picture oversimplifies the problem but nevertheless provides a basis for understanding the rotational stabilization of external modes that grow on the $\tau_W$ time scale, regardless of whether it is the plasma or the wall that is moving.

Unfortunately, a rotating wall will also tend to affect the equilibrium fields. If $|\mathbf{k}_{eq} \cdot \mathbf{u}| \gg \Omega_W$, where the wave vector $\mathbf{k_{eq}}$ now measures the spatial variation of these fields, they will be either "frozen-in" the liquid metal and get dragged by it, if they have already penetrated the fluid, or they will be shielded, like the perturbed fields of an unstable mode. Both scenarios will generate large volume and surface currents in the liquid metal. In order not to

perturb the equilibrium fields, the flow would have to be an equilibrium flow of the form[21]

$$\mathbf{u_{eq}} = \frac{F(\psi)}{\rho}\mathbf{B} + R^2\Omega(\psi)\nabla\zeta, \qquad (7)$$

where $F(\psi)$, and $\Omega(\psi)$ are arbitrary flux functions, $\rho$ is the mass density, and $\zeta$ is the toroidal angle. Note that $\mathbf{u_{eq}}$ can be purely toroidal but cannot be purely poloidal, as contemplated by the proponents of the liquid metal wall concept. Consequences of violating the equilibrium condition are graphically demonstrated in Figure 7. In Fig. (7a) the pressure contours in an elongated, n=0 unstable equilibrium, depict the induced rotation in the plasma when the liquid wall starts rotating poloidally (but not within flux surfaces) with rotation frequency $\Omega_R = 0.2\Omega_W$ (for $k = 1/a$), due to induced co-rotating toroidal currents in the liquid. Even when a poloidal flow is almost entirely within flux surfaces, as in Fig. 7b, it leads to large poloidal currents in the liquid wall, $\mathbf{J}_{pol} \geq 10\ \mathbf{J}_{eq}$ for $\Omega_R/\Omega_W \sim \mathcal{O}(1)$ due to dragging of the toroidal field. These currents would be expected to interfere and stop the flow.

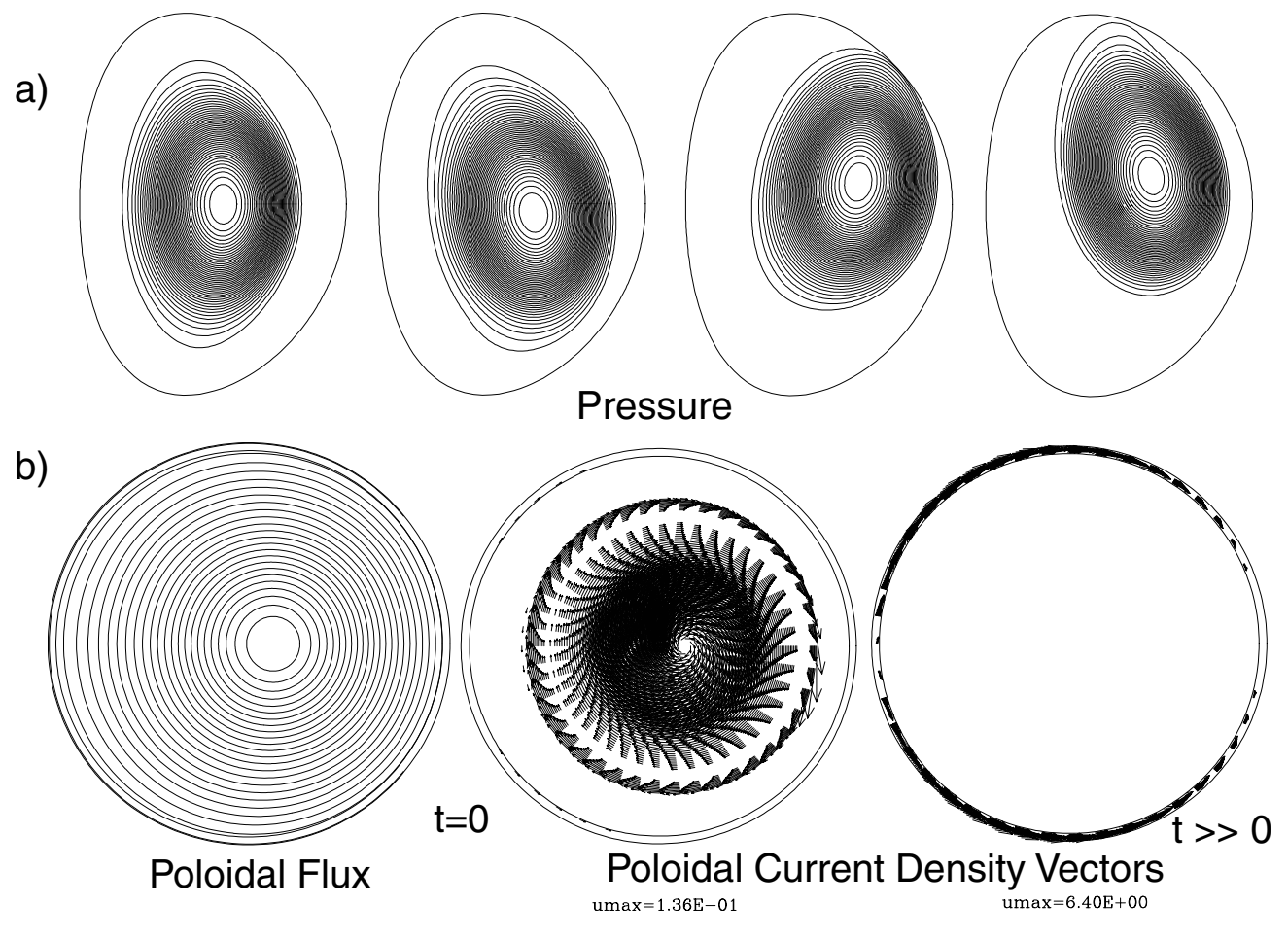

FIG. 7: Effects of uniformly rotating "liquid metal wall", modelled by a moving conductor, on an elongated (a), and circular (b) equilibrium.

Unless a technologically feasible method can be found to rotate the liquid metal with an equilibrium velocity given above, basing advanced reactor designs on the liquid metal wall concept seems rather ill-advised at this point.

This work was supported by the U.S. DoE under grant No. DE-FG03-96ER-54346.